\documentclass[sigconf,anonymous=false,nonacm]{acmart}

\AtBeginDocument{%
  \providecommand\BibTeX{{%
    \normalfont B\kern-0.5em{\scshape i\kern-0.25em b}\kern-0.8em\TeX}}}

\begin{document}

\title[Systolic-CNN: An OpenCL-defined Scalable Run-time-flexible FPGA Accelerator Architecture]{Systolic-CNN: An OpenCL-defined Scalable Run-time-flexible FPGA Accelerator Architecture for Accelerating Convolutional Neural Network Inference in Cloud/Edge Computing}

\author{Akshay Dua}
\affiliation{\institution{Arizona State University}}
\email{adua5@asu.edu}

\author{Yixing Li}
\affiliation{\institution{Arizona State University}}
\email{yixingli@asu.edu}

\author{Fengbo Ren}
\affiliation{\institution{Arizona State University}}
\email{renfengbo@asu.edu}



\begin{abstract}
This paper presents Systolic-CNN, an OpenCL-defined scalable, run-time-flexible FPGA accelerator architecture, optimized for accelerating the inference of various convolutional neural networks (CNNs) in multi-tenancy cloud/edge computing. The existing OpenCL-defined FPGA accelerators for CNN inference are insufficient due to limited flexibility for supporting multiple CNN models at run time and poor scalability resulting in underutilized FPGA resources and limited computational parallelism. Systolic-CNN adopts a highly pipelined and paralleled 1-D systolic array architecture, which efficiently explores both spatial and temporal parallelism for accelerating CNN inference on FPGAs. Systolic-CNN is highly scalable and parameterized, which can be easily adapted by users to achieve up to 100\% utilization of the coarse-grained computation resources (i.e., DSP blocks) for a given FPGA. Systolic-CNN is also run-time-flexible in the context of multi-tenancy cloud/edge computing, which can be time-shared to accelerate a variety of CNN models at run time without the need of recompiling the FPGA kernel hardware nor reprogramming the FPGA. The experiment results based on an Intel Arria/Stratix 10 GX FPGA Development board show that the optimized single-precision implementation of Systolic-CNN can achieve an average inference latency of 7ms/2ms, 84ms/33ms, 202ms/73ms, 1615ms/873ms, and 900ms/498ms per image for accelerating AlexNet, ResNet-50, ResNet-152, RetinaNet, and Light-weight RetinaNet, respectively. Codes are available at https://github.com/PSCLab-ASU/Systolic-CNN.
\end{abstract}

\ccsdesc[500]{Hardware~Hardware accelerators}
\ccsdesc[300]{Computer systems organization~Neural networks}

\keywords{FPGA, neural networks, OpenCL, accelerator}



\maketitle

\section{Introduction}


FPGAs offer superior hardware flexibility and energy efficiency that have attracted many researchers and developers to use FPGAs for accelerating convolutional neural network (CNN) inference for computer vision tasks \cite{ref11_autoSystolic,ref15_pipecnn,ref16_throughput-opt}. The conventional development flow of FPGAs relies on designing FPGA hardware at the register-transfer level (RTL). Although it allows the fine control of resource utilization for precise performance improvement \cite{ref10_CNNfixedpoint}, the large efforts needed in design and verification make architecture design space exploration time-consuming. High-level synthesis (HLS) tools, such as the Intel FPGA SDK for OpenCL, allow function modeling at a much higher level, thus enabling a faster design and verification cycle. The HLS tools also provide a rich set of synthesis attributes and directives that facilitates efficient architecture design space exploration \cite{advantagesHLS}. 

Many recent works explore accelerating CNNs on FPGAs using C/OpenCL showing promising acceleration performance \cite{ref9,ref11_autoSystolic,ref12_YOLO,ref15_pipecnn,ref16_throughput-opt,ref11_autoSystolic,ref17_freqImproveSystolic}. Nevertheless, these works suffer from two major limitations that make them insufficient for realizing acceleration-as-a-service for multi-tenancy cloud/edge computing: 1) the lack of flexibility for supporting multiple CNN models at run time; and 2) the poor scalability resulting in underutilized FPGA resources and limited computational parallelism. 

In this paper, we present Systolic-CNN, an OpenCL-defined scalable, run-time-flexible FPGA accelerator architecture for accelerating CNN inference in multi-tenancy cloud/edge computing. Systolic-CNN adopts a highly pipelined and paralleled 1-D systolic array architecture, which efficiently explores both spatial and temporal parallelism for accelerating CNN inference on FPGAs. Systolic-CNN is highly scalable and has three key architectural parameters, based on which a user can optimally scale the accelerator architecture to fully utilize the off-chip memory bandwidth and available DSP block resource given an FPGA board. In addition, single-precision Systolic-CNN is run-time-flexible in the context of multi-tenancy cloud/edge computing, which can be time-shared to accelerate a variety of CNN models (including error-sensitive applications \cite{IRthermography,defectDetection}) at run time without the need of recompiling the FPGA kernel hardware nor reprogramming the FPGA. The experiment results based on an Intel Arria/Stratix 10 GX FPGA Development board show that the optimized single-precision implementation of Systolic-CNN can achieve an average inference latency of 7ms/2ms, 84ms/33ms, 202ms/73ms, 1615ms/873ms, and 900ms/498ms per image for accelerating AlexNet, ResNet-50, ResNet-152, RetinaNet, and Light-weight RetinaNet, respectively. The peak computational throughput is measured at 80-210 GFLOPS/s and 242-700 GFLOPS/s for accelerating different single-precision CNN models on Arria/Stratix 10 FPGA board. The source codes are available at https://github.com/PSCLab-ASU/Systolic-CNN.


\section{Background}

\subsection{CNN Models}
\begin{figure}[b]
    \centering
    \includegraphics[width=7cm]{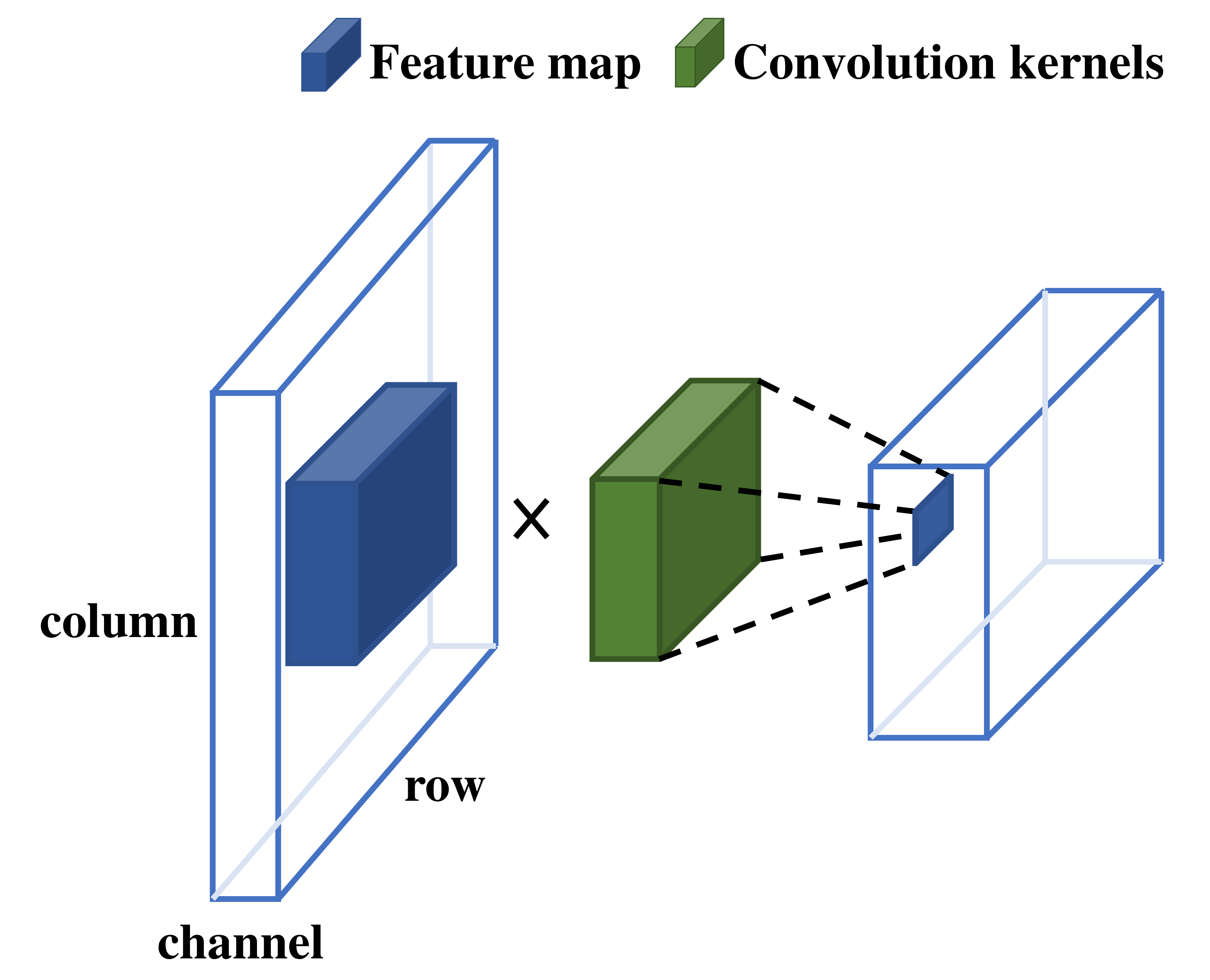}
    \caption{A convolutional layer.}
    \label{fig: conv}
\end{figure}

CNNs are a subcategory of deep neural network models. CNNs can extract and learn spatial information automatically to process image classification and object detection tasks \cite{resnet,retinanet,lwretinanet}. CNNs typically consist of convolutional layers, fully-connected layers, pooling layers and non-linearity layers \cite{dlnature}. Among them, convolutional layers consume most of the computations. A standard convolutional layer is shown in Fig. \ref{fig: conv} \cite{angeleye}. The convolutional kernels slide over the input feature maps and compute the inner-products for the output feature maps. It's possible to have both spatial and temporal parallelism to perform high-throughput convolutions. 

In this paper, we evaluate the performance of the proposed accelerator architecture on five different CNN models, namely AlexNet \cite{alexnet}, ResNet-50 \cite{resnet}, ResNet-152 \cite{resnet}, RetinaNet \cite{retinanet}, and lightweight (LW) RetinaNet \cite{lwretinanet}. The first three CNN models are used for classification tasks, while the other two are used for object detection tasks. 

\subsection{Related Works on OpenCL-defined FPGA Accelerators for CNN Inference}
There have been many works \cite{ref9,ref11_autoSystolic,ref12_YOLO,ref15_pipecnn,ref16_throughput-opt,ref17_freqImproveSystolic} on accelerating CNNs on FPGAs using C/OpenCL published in recent years. However, these works suffer from one major limitation that makes them insufficient for handling the dynamic workloads in a multi-tenancy cloud/edge computing environment. Most of the existing works are either designed to exclusively accelerate a specific CNN model \cite{ref12_YOLO} or requires recompilation of the FPGA kernel and reprogramming of the FPGA device when changing the CNN model for acceleration \cite{ref11_autoSystolic,ref16_throughput-opt, ref15_pipecnn}. For example, the work in \cite{ref12_YOLO} shows a high inference throughput but is restricted to accelerating a YOLO CNN model only \cite{redmon2017yolo9000}. PipeCNN \cite{ref15_pipecnn}, although designed for accelerating a variety of CNN models, requires updating the line buffer size and recompiling the FPGA kernel code for each CNN model due to the folded computation of inner product along the channel dimension that varies upon different CNN models. Given that FPGA kernel compilation can take a long time, this is a deal-breaker for providing acceleration-as-a-service in multi-tenancy cloud/edge computing. Differently, the architecture design of Systolic-CNN is completely invariant to CNN models. Specifically, the memory access pattern of input feature maps (IFMs) and local buffer sizes in Systolic-CNN only depend on user-defined architecture parameters regardless of the CNN model mapped. Thus, Systolic-CNN is run-time-flexible, which can be time-shared to accelerate a variety of CNN models at run time without the need to recompile the FPGA kernel hardware nor reprogramming the FPGA.

In addition, many of the existing works have poor scalability \cite{ref11_autoSystolic,ref15_pipecnn,ref16_throughput-opt}, which can be evidenced by the reported under-utilization of the coarse-grained computing resources available on an FPGA ($<90\%$ DSP block utilization). For example, although PipeCNN \cite{ref15_pipecnn} exploits two levels of spatial parallelism, we notice in our experiment that scaling up the computation parallelism (e.g. $vec\_size$) in PipeCNN \cite{ref15_pipecnn} can create difficulty for the placement-and-routing stage due to the large fan-out required at local memory buffer interfaces. \cite{ref11_autoSystolic,ref17_freqImproveSystolic} propose to adopt a 2-D systolic array architecture \cite{ref18_vlsiProcessor} to improve the design scalability and operating frequency but still fail to fully utilize the available DSP block resources as the optimal mapping of a 2-D systolic array highly depends on the physical layout of an FPGA, which can change across different FPGA boards. Differently, Systolic-CNN adopts a highly pipelined and paralleled 1-D systolic array architecture with shift-register based IFM buffers, which efficiently explores both spatial and temporal parallelism as well as the data reuse of IFMs to improve inference throughput with reduced off-chip memory access and well-bounded fan-out. As a result, Systolic-CNN is highly scalable and parameterized, which can be easily adapted by users to achieve up to 100\% utilization of the coarse-grained computation resources (i.e., DSP blocks) for a given FPGA. 
\section{Architecture Design}

\begin{figure}[b]
    \centering
    \includegraphics[width=8cm]{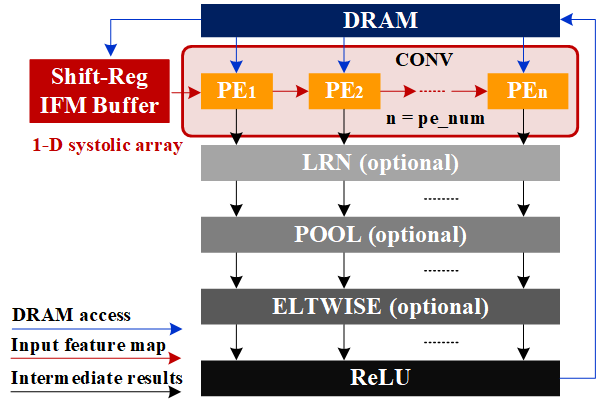}
    \vspace*{-4mm}
    \caption{The system architecture of Systolic-CNN.}
    \label{fig: design architecture}
\end{figure}

\subsection{System Architecture}
Fig. \ref{fig: design architecture} shows the high-level system architecture of Systolic-CNN. The convolution engine (CONV) in Systolic-CNN adopts a highly pipelined and paralleled 1-D systolic processing element (PE) array architecture \cite{kung1982systolic} for performing high-throughput convolutions with both spatial and temporal parallelism. The IFMs are read from the off-chip memory and cached in an on-chip shift-register-based IFM buffer for reducing off-chip memory access and maximizing the data reuse of IFMs for the convolution computation both within the same and across different output feature maps (OFMs). The weights are also read from the off-chip memory and cached inside the PEs to be reused for the convolution computation within the same OFMs. The system architecture also implements other commonly used layers in CNN models, namely normalization layer (LRN), element-wise sum layer (ELTWISE), rectified linear unit layer (ReLU), and pooling layer (POOL). The LRN, POOL, ELTWISE, and ReLU computation are optional during the kernel execution, depending on the CNN model structure. The final output results are loaded back to the off-chip memory for either the next round of computation or return to the host kernel process.

To allow Intel FPGA SDK for OpenCL to better resolve the data-dependency and create the deep processing pipeline properly, the FPGA kernels of Systolic-CNN are all implemented as single-threaded kernels. Specifically, the shift-register-based IFM buffer is implemented as the MemRD kernel, each PE is implemented as an auto-run kernel to minimize the host-induced latency during CNN inference, the LRN and POOL modules are each implemented as a separate kernel, and the ELTWISE and ReLU modules are combined and implemented as the MemWrite kernel. Given that convolution is the bottleneck of computation in CNNs, the PE (convolution) kernels are designed to utilize most of the coarse-grained computation resources on an FPGA, while the other computation kernels are designed to utilize the minimum resources needed for making sure they are not the computational throughput bottleneck. In addition, the PE (convolution) kernels are optimized with a minimum initiation interval of 1 cycle. Systolic-CNN can support any customized residual neural networks with skipped connections. 

\subsection{Architectural Parameters}
We parameterize the system architecture of Systolic-CNN with three architectural parameters, namely $pe\_num$, $vec\_fac$, and \\$reuse\_fac$. 

From a system architecture perspective, $pe\_num$ defines the number of PEs in the 1-D systolic array that performs temporally paralleled convolution in a deep pipeline. Each PE performs the convolution computation of a different OFM by sharing the same IFM data in a shifted fashion. Thus, $pe\_num$ also defines the parallelism of OFM generation. $reuse\_fac$ defines the parallelism of the inner product (IP) units inside each PE as well as how many times the same IFM data is reused by each PE for the convolution computation within the same OFM. Increasing $reuse\_fac$ will improve the computational throughput without changing the amount of off-chip memory access needed for reading the IFMs, thus relaxes the off-chip memory bandwidth requirement and improves the off-chip memory bandwidth efficiency. $vec\_fac$ defines the SIMD width of the partial IP computation between the weight vector and IFM vector across $vec\_fac$ different channels inside each IP unit in each PE. Thus, $vec\_fac$ and $reuse\_fac$ also defines the parallelism of IFM computation along the channel and the row dimension of the IFMs, respectively. In addition, the size of the shift-register-based IFM buffer is defined by $reuse\_fac \times vec\_fac$. These three parameters allow users to efficiently perform architecture design space exploration to maximize the resource utilization of a given FPGA board subject to the available off-chip memory bandwidth. An example of the design space exploration is discussed in Section 4.2. 

\begin{figure}[tb]
    \includegraphics[width=7cm]{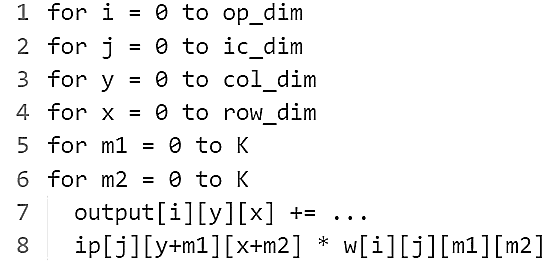}
    \vspace*{-2mm}
    \caption{The pseudo code for a standard convolutional layer}
    \label{fig: code1}
\end{figure}

\begin{figure}[b]
    \includegraphics[width=7.5cm]{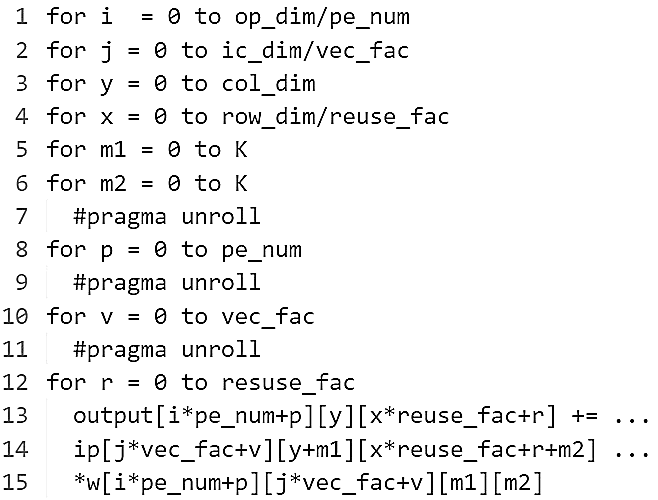}
    \caption{The pseudo code for a convolutional layer with the three architectural parameters implemented.}
    \label{fig: code2}
\end{figure}

To illustrate the impact of the architectural parameters on acceleration performance, we provide the pseudo code of a standard convolutional layer in Fig. \ref{fig: code1}, and a convolutional layer optimized with the three architectural parameters implemented in Fig. \ref{fig: code2}. From an algorithmic perspective, $pe\_num$, $vec\_fac$, and $reuse\_fac$ can be interpreted as the unrolling factor of the for loop along the depth of OFM ($op\_dim$), the depth (channel dimension) of IFM ($ic\_dim$), and the row dimension of the IFM ($row\_dim$), respectively. It should be noted that the system architecture of Systolic-CNN only depends on the three architectural parameters that are completely invariant to CNN models. Such invariance is the key to enabling the run-time flexibility needed for handling the dynamic workload in a multi-tenancy cloud/edge computing environment. 

\begin{figure*}[!tb]
    \centering
    \includegraphics[width=13cm]{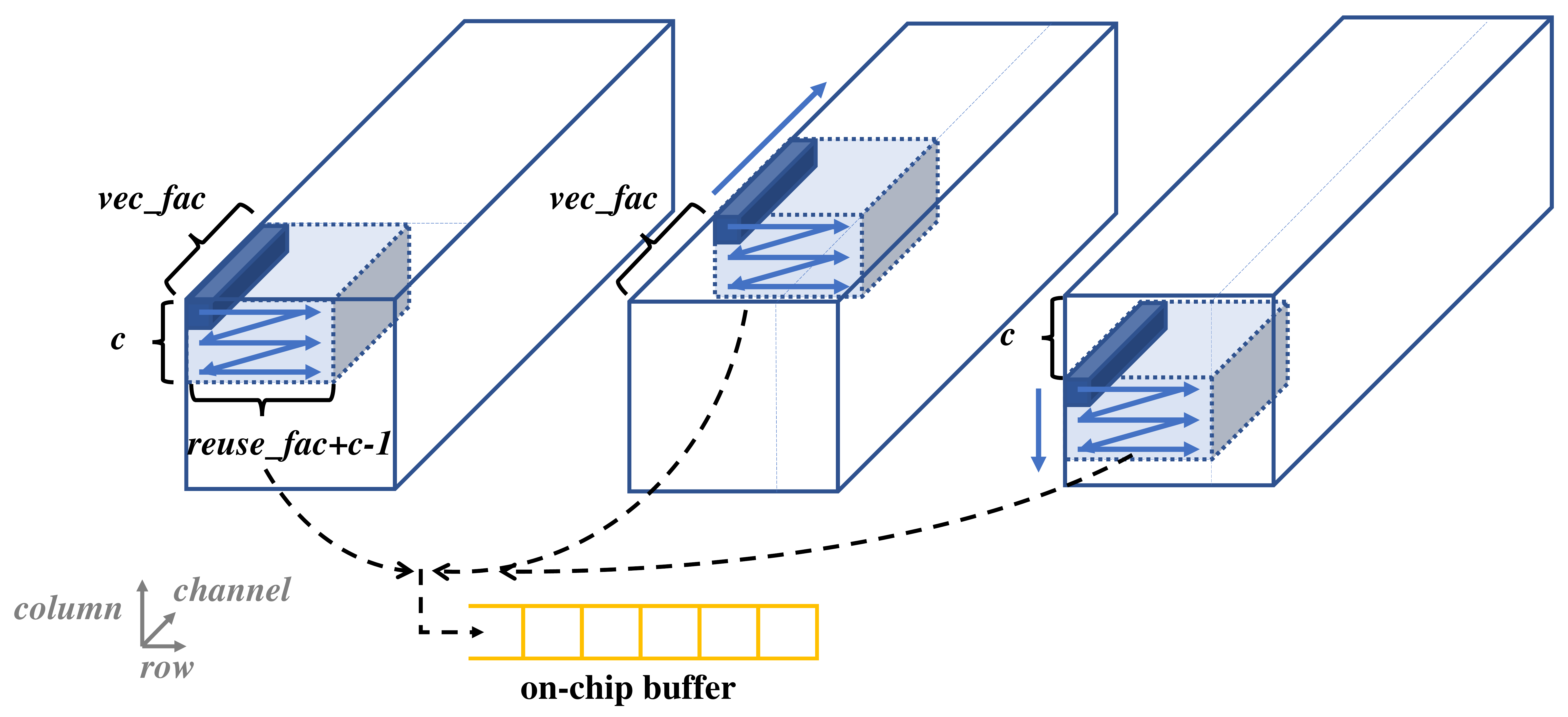}
    \caption{Data loading scheme of the input feature map.}
    \label{fig: IFM}
\end{figure*}

\subsection{Data loading scheme}
Fig. \ref{fig: IFM} shows the data loading scheme of the input feature map (IFM). In one clock cycle, $1 \times 1 \times vec\_fac$ IFM values (highlighted in dark blue) are loaded onto shift-register-based IFM buffer. Then, the loading window slides along the row dimension $reuse\_fac + c - 1$ times and slides along the column dimension $c$ times, where $c$ represents the kernel size of a convolution kernel. Thus the IFM values can be reused $reuse\_fac$ times computing with the $c \times c $ convolution kernel. After the buffered IFM values have done all the computations with different weights, the loading window slides over the channel dimension to repeat the operations stated above.

\subsection{PE Design}

\begin{figure*}[!tb]
    \centering
    \includegraphics[width=15cm]{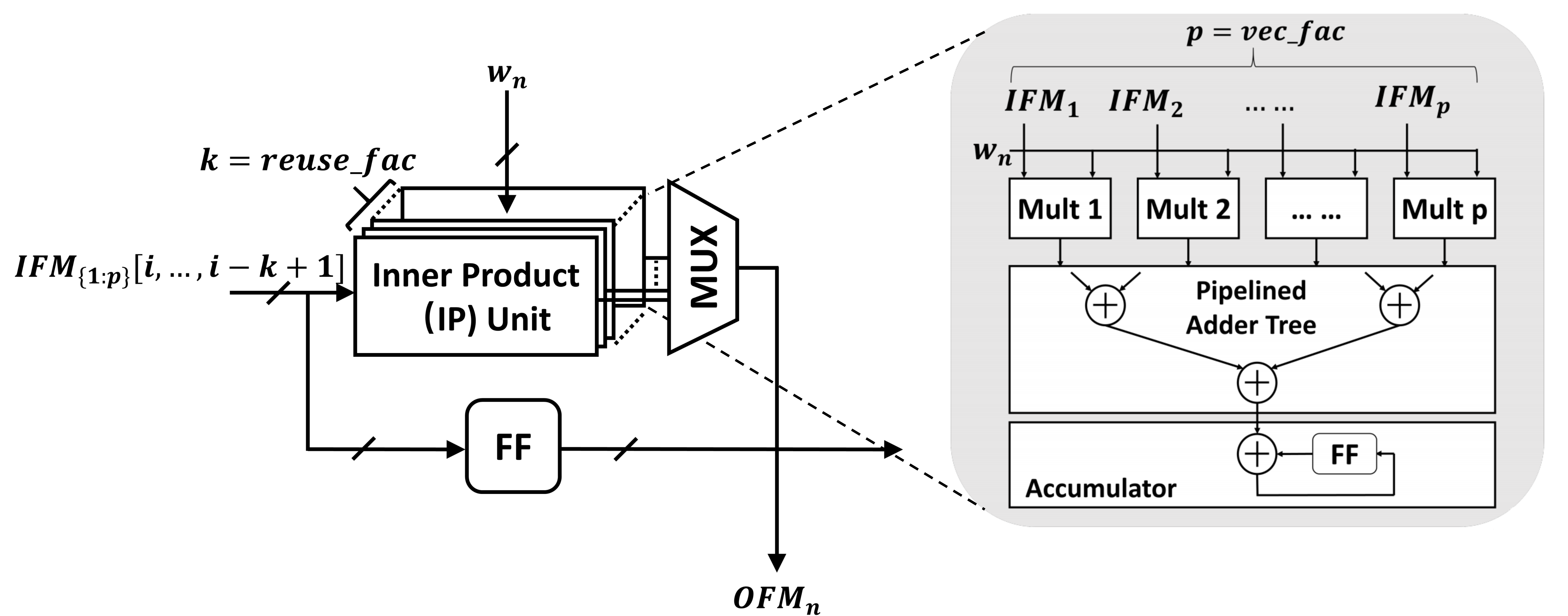}
    \caption{The architecture of the n\textsuperscript{th} PE.}
    \label{fig: PE}
\end{figure*}

Fig. \ref{fig: PE} illustrates the architecture of the n\textsuperscript{th} PE in the convolution engine of Systolic-CNN. Each PE contains multiple IP units (defined by $reuse\_fac$), each of which computes the 3D inner product across different sliding windows of the convolution computation within the same OFM. Different IP units share the same set of weights and take in the same IFM vector sequence in a shifted fashion to reuse the IFM data by a factor of $reuse\_fac$ times. Each IP unit contains multiple multipliers and a pipelined adder tree for computing partial IPs with a SIMD width defined by $vec\_fac$ as well as an accumulator for computing the IP of an arbitrary dimension in a folded, pipelined fashion to eliminate the need of data movement for partial IP summation. To facilitate the IFM data movement throughout the 1D systolic array of PEs, each PE also shifts the input IFM data directly to the subsequent PE with a one cycle latency. 

It should be noted that when performing the computation in fully connected layers, the weight sharing across different IP units in the PE becomes inefficient and causes low utilization of the computation resources. To address this problem, Systolic-CNN supports a batch processing mode for fully connected layer computation. By processing multiple input images in a batch mode, the same weights in the fully connected layer can be again shared across different IP units for performing the computation of different images. The batch size must be $\leq reuse\_fac$. When the batch size $=reuse\_fac$, the computation resources in each PE can be fully utilized for accelerating the computation in fully connected layers.

The convolution computation performed in each PE exploits two levels of spatial parallelism: the parallelism of IP units defined by $reuse\_fac$ and the parallelism of partial IP computation defined by $vec\_fac$. Given the 1D systolic PE array also exploit a temporal/pipelined parallelism of $pe\_num$, the overall parallelism of convolution computation employed in Systolic-CNN is $vec\_fac \times reuse\_fac \times pe\_num$. 

Based upon the understanding of the PE architecture, one should note that while increasing any of the three architectural parameters keeps the promise to improve the computation parallelism and the computational throughput proportionally, their impact on the required off-chip memory bandwidth is slightly different. Increasing $vec\_fac$ increases the amount of IFM data accessed in each clock cycle, thus has a large impact on the required off-chip memory bandwidth. Increasing $pe\_num$ increases the amount of weight data access required only at the beginning of each convolution computation or in each clock cycle in the case of fully connected layer computation. Thus, $pe\_num$ has a large impact on the required off-chip memory bandwidth during the computation of fully connected layers. Differently, increasing $reuse\_fac$ will only change IFM data access pattern without affecting the amount of IFM data accessed in each clock cycle, thus have no impact on the required off-chip memory bandwidth.

The advantages of the 1-D systolic PE array architecture include 1) limiting the fan-out at the local IFM buffer interface; 2) assuring short and local interconnects used in the FPGA implementation; 3) reducing the amount of off-chip and on-chip memory access needed by reusing and moving IFM data through shift registers. These benefits are the key to improving the scalability of Systolic-CNN, the system operating frequency and the off-chip memory bandwidth efficiency, which are all essential to the system-level performance of CNN acceleration on an FPGA computing device. Compared with 2-D systolic array-based CNN accelerator architectures \cite{ref11_autoSystolic,ref17_freqImproveSystolic}, the 1-D systolic PE array architecture of Systolic-CNN has much more simplified memory control, data organization, and local buffering schemes for handling IFM and weight data.


\subsection{Design for Scalability}
The proposed 1D systolic PE array architecture of Systolic-CNN resolves the routing congestion problem caused by the large fan-out issue at the local memory buffer interface that exists in the current work \cite{ref15_pipecnn} that exploits spatial parallelism only using NDRange kernels. However, as the design scales up, we observe that the off-chip memory controller, i.e. the load-store unit (LSU), automatically synthesized by Intel FPGA SDK for OpenCL starts to show a large fan-out, which becomes the new bottleneck of routing congestion that prevents the design from further scaling up. 

The high fan-out issue exists when the value of either $vec\_fac$ and $pe\_num$ is high. When $vec\_fac$ and $pe\_num$ is high, a large fan-out is observed at the LSU interface for loading the IFMs and the weights in parallel, respectively. Knowing that $vec\_fac$ has a much bigger impact on the required off-chip memory bandwidth than $pe\_num$ during convolution computation, one should consider limiting the value of $vec\_fac$ to avoid a memory-bounded design regardless. Thus, $pe\_num$ and the parallel loading of weights that are more likely to be the problems here. To resolve the high fan-out issue of the LSU, we propose to generate multiple LSUs to transfer the weights from the off-chip memory to local buffers in a sequential manner instead. In the case of $pe\_num=16$, we observe that the proposed solution not only resolves the routing congestion problem but also improves the system operating frequency by 10$\%$. This is the key to allowing users to further scale up the design to efficiently utilize the DSP blocks on an FPGA.

\subsection{Host Kernel Design}
While the FPGA kernels of Systolic-CNN are invariant to CNN models, a host kernel must be customized for deploying different CNN models onto the Systolic-CNN implementation on an OpenCL-supported FPGA computing device. The host kernel should invoke the corresponding computation kernel in Systolic-CNN just once for mapping each layer of a CNN model depending on the CNN model structure. The CNN model parameters (filter sizes, stride, padding information, etc.) are sent from the host kernel program to the FPGA kernels at run time to control the operations of each of the invoked FPGA kernel. The run-time flexibility of Systolic-CNN allows edge users to deploy a wide range of CNN models for acceleration without the need to change or recompile the FPGA kernel codes nor reprogramming the FPGAs. This is the key to enabling the acceleration-as-a-service for CNN inference in multi-tenancy cloud/edge computing.

\section{Experiments}

\subsection{Experimental Setup}
We use two different settings to conduct experiments for edge and cloud computing scenarios. The experiments to reflect the edge computing user cases are conducted based on an Intel Arria 10 GX FPGA Development board that is equipped with an Intel 10AX115S2F45I1SG FPGA and 2GB DDR4 SDRAM with a maximum memory bandwidth of 19.2 GB/s. We use Intel FPGA SDK for OpenCL version Pro 18.0 for kernel compilation and deployment. The experiments to reflect the cloud computing user cases are conducted based on BittWare 520N FPGA accelerator card that is equipped with a Stratix 10 GX2800 FPGA and 32GB DDR4 SDRAM with a maximum memory bandwidth of 2400 MT/s. We use Intel FPGA SDK for OpenCL version Pro 19.4 for kernel compilation and deployment. 

Systolic-CNN adopts the single-precision floating-point data format for the sake of run-time flexibility\textemdash to maintain a sufficiently large dynamic range for supporting different CNN models (including error-sensitive applications, such as industrial robots or medical-related applications \cite{IRthermography,defectDetection}) at run time. 

\subsection{Design Space Exploration}
A key design target of Systolic-CNN is to efficiently utilize the available DSP resources on an FPGA to maximize the computation parallelism and computational throughput for CNN inference subject to the available off-chip memory bandwidth. In this section, we use the AlexNet as an example to demonstrate the space exploration of the three architectural parameters with respect to the Intel Arria 10 GX FPGA Development board. Given the different impact on the off-chip memory bandwidth requirement, the values of the three architectural parameters shall be determined in the order of 1) $vec\_fac$, 2) $pe\_num$, and 3) $reuse\_fac$.




\subsubsection{$vec\_fac$}
$vec\_fac$ determines the parallelism of IFM data access from the off-chip memory to the shift-register-based IFM buffer per clock cycle, thus has a large impact on the off-chip memory bandwidth. As a result, the value of $vec\_fac$ should depend on the per-cycle burst width of data access allowed by the off-chip memory and the bit width of the IFM. Specifically, the optimal value of $vec\_fac$ can be calculated as $vec\_fac = burstWidth/bitWidth$.
Given the value of $vec\_fac$ determined by the equation above, there will be no memory stalling when the off-chip memory access of IFM data happens every clock cycle (the convolution kernels operate with a minimum initiation interval of 1 cycle), which guarantees a high off-chip memory bandwidth efficiency. Since the burst width of data access allowed by the off-chip memory on the Intel Arria 10 GX FPGA Development board is 512 bits and the bit width of IFM data is 32 bits based on the single-precision floating-point data format, we set the value of $vec\_fac$ to 16 in our experiments. 

\subsubsection{$pe\_num$}
\begin{figure}[tb]
    \centering
    \includegraphics[width=8.5cm]{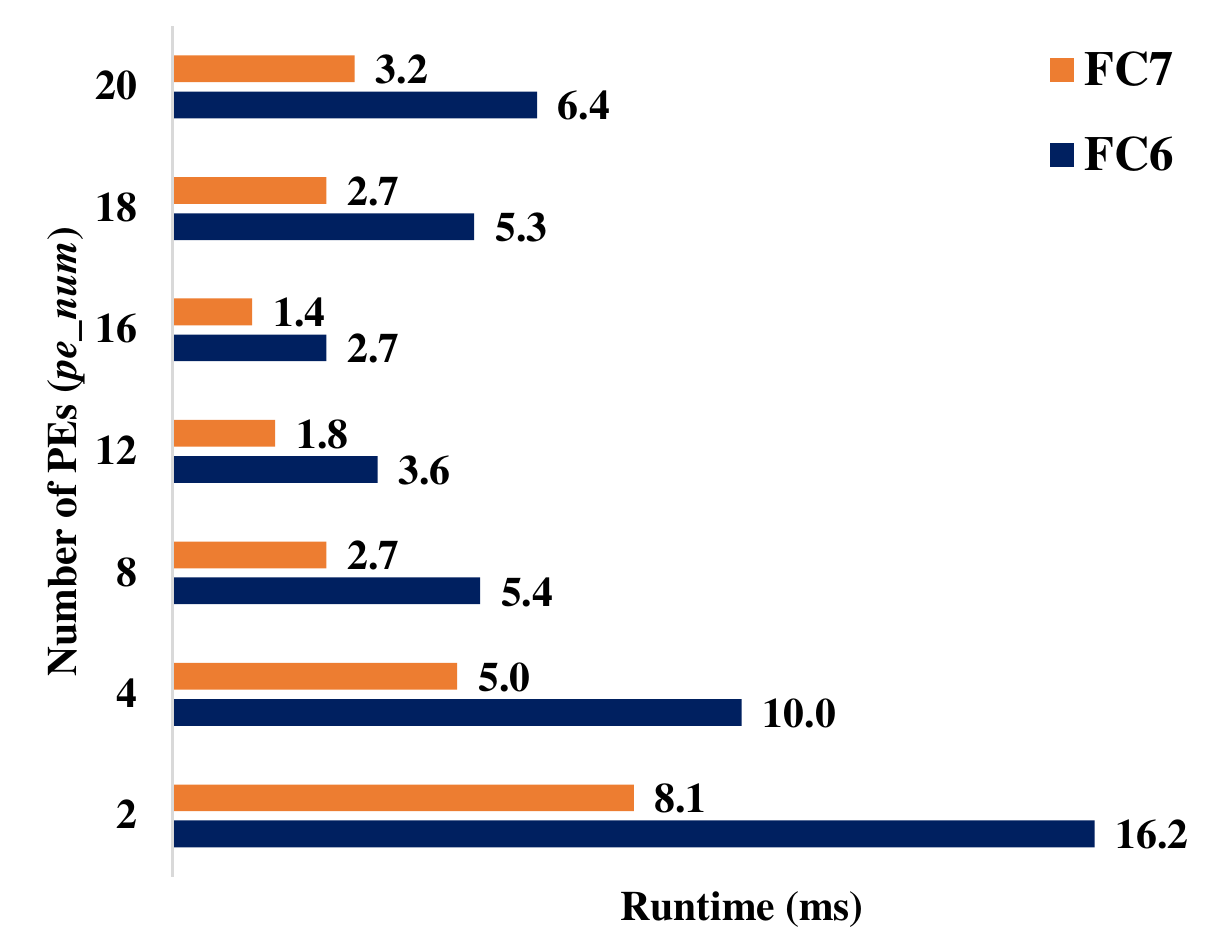}
    \vspace*{-2mm}
    \caption{Runtime (ms) of the FC6 and FC7 layers in AlexNet at different $pe\_num$ ($vec\_fac$=16, $reuse\_fac$=1 ) on Arria 10 FPGA board.}
    \label{fig:alexnet2}
\end{figure}

$pe\_num$ determines the parallelism of weight data access from the off-chip memory per clock cycle for fully connected layer computation thus has a large impact on the off-chip memory bandwidth during the computation of fully connected layers only. To determine the optimized value of $pe\_num$, we measure the runtime of the top two memory-intensive layers in the AlexNet -- FC6 and FC7 (FC stands for a fully-connected layer) at different $pe\_num$. As shown in Fig. \ref{fig:alexnet2}, $pe\_num$ is swept from 2 to 20 with a step size of 2, while $vec\_fac$ is fixed to 16 and $reuse\_fac$ is set to 1. The runtime of FC6 and FC7 layers reaches the minimum at the $pe\_num$ value of 16. The increase in runtime beyond the $pe\_num$ value of 16 indicates that those cases are already memory-bounded, which are limited by the available off-chip memory bandwidth. Therefore, the optimal value of $pe\_num$ is determined to be 16 in our experiments.

\begin{figure}[tb]
    \centering
    \includegraphics[width=8.5cm]{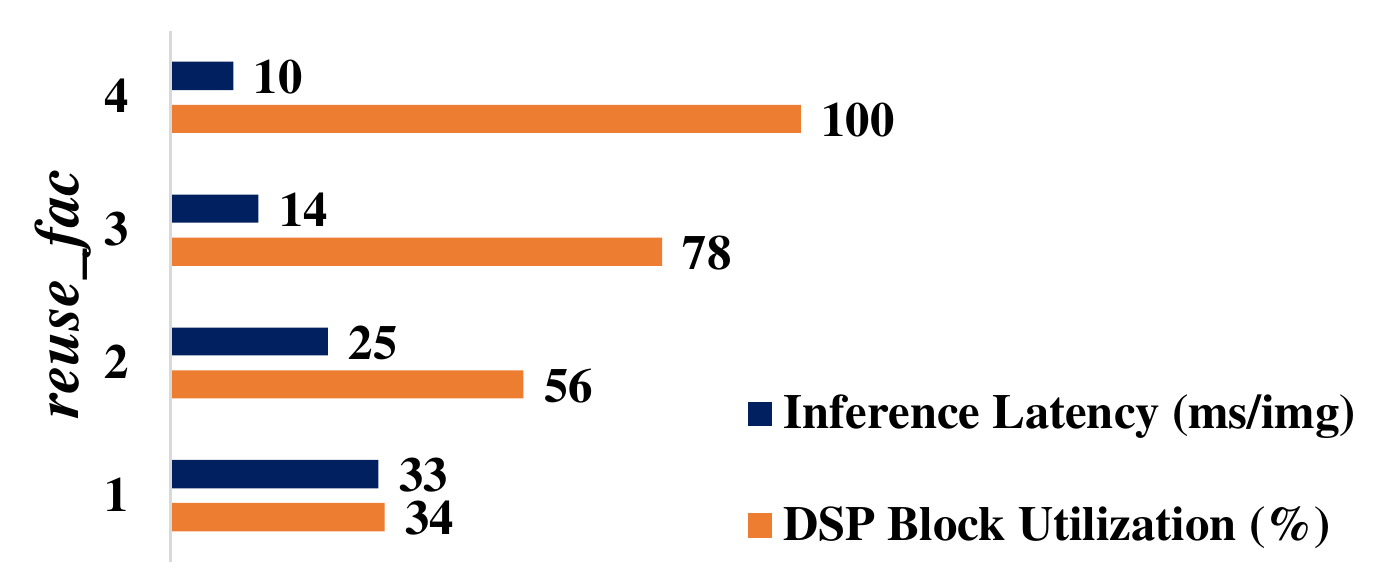}
    \caption{Inference latency (ms/img) of AlexNet and DSP block utilization (\%) at different $reuse\_fac$ ($vec\_fac$=16, $pe\_num$=16) on Arria 10 FPGA board.}
    \label{fig:alexnet1}
\end{figure}

\subsubsection{$reuse\_fac$}
$reuse\_fac$ determines the parallelism of IP units inside each PE for reusing the IFM data as well as the size of the shift-register-based IFM buffer. Since $reuse\_fac$ has no impact on the off-chip memory bandwidth requirement, the scaling of $reuse\_fac$ is not limited by the off-chip memory characteristics but only depends on the available DSP resources on an FPGA. Fig. \ref{fig:alexnet1} shows the inference latency of running the entire AlexNet \cite{alexnet} on the ImageNet dataset \cite{deng2009imagenet} and the DSP block utilization at different $reuse\_fac$, which is swept from 1 to 4, while $vec\_fac$ and $pe\_num$ are both fixed to 16. It is shown that the DSP utilization increases and the runtime decrease both in a linear fashion as $reuse\_fac$ increases. In addition, the DSP utilization of 100\% and the minimum runtime are achieved at the $reuse\_fac$ value of 4. The results shown in Fig. \ref{fig:alexnet1} illustrate the great scalability of Systolic-CNN. By optimizing the three architectural parameters of Systolic-CNN following the guidelines, one can efficiently utilize the available FPGA resources to maximize the computational throughput of CNN inference subject to the available off-chip memory bandwidth. 

Through the design space exploration, the optimal value of $pe\_num$, $reuse\_fac$ and $vec\_fac$ with respect to the Intel Arria 10 GX FPGA Development board is found to be 16, 4 and 16, respectively. We use the same methodology to explore three architectural parameters for Intel Stratix 10 GX FPGA Development board. The optimal value of $pe\_num$, $reuse\_fac$ and $vec\_fac$ is found to be 16, 6, 32, respectively.

\subsection{Experiment Results}
We measure the inference latency of the optimized Systolic-CNN accelerators on an Intel Arria 10 GX FPGA Development board and an Intel Stratix 10 GX2800 FPGA Development board for running five different CNN models: AlexNet \cite{alexnet}, ResNet-50 \cite{resnet}, ResNet-152 \cite{resnet}, RetinaNet \cite{retinanet}, and LW-RetinaNet \cite{lwretinanet}, respectively. As Systolic-CNN is run-time-flexible, only the host kernel is updated for deploying different CNN models on a single board without recompiling or redeploying the FPGA kernel. '

The purpose of the comparison with state-of-the-art is not to show any performance benefits of Systolic-CNN, but rather, it is to show the runtime flexibility and scalability advantages with the comparable performance given the differences in data format, numerical precision, and computational methods used in different designs.


 \begin{table*}[tb]
    \centering
    \caption{{Comparison with Prior OpenCL-based FPGA Accelerators for AlexNet.}}
    \begin{tabular}{|p{80pt}|p{40pt}|p{40pt}|p{40pt}|p{50pt}|p{60pt}|}
    \hline
    Work & \cite{ref9} & \cite{ref11_autoSystolic} & \cite{ref15_pipecnn} & \cite{ref16_throughput-opt} & This work\\
    \hline
    FPGA & Arria 10 GT1150 & Arria 10 GT1150 & Arria 10 GX1150 & Stratix-V GSD8 & Arria 10 GX1150\\
    \hline
    CNN Model & AlexNet & AlexNet & AlexNet & AlexNet & AlexNet\\
    \hline
    Data Format & 16-bit float &  32-bit float & 8-bit fixed & 8/16-bit fixed & 32-bit float\\
    \hline
    Logic Utilization & 246K (58\%) & 350K (82\%) & 105K (25\%) & N/A & 250K (59\%)\\
    \hline
    Memory Utilization & 2487 (92\%) & 2360 (86\%) & 641 (24\%) & N/A & 2472 (91\%)\\
    \hline
    DSP Utilization & 1476 (97\%) & 1290 (85\%) & 377 (25\%) & N/A & 1518 (100\%)\\
    \hline
    Inference Latency & 1ms & 4ms & 22ms & 20ms & 10ms/7ms (non-batch/batch)\\
    \hline
    $f_{CLK}$ & 303MHz & 239MHz & 250MHz & 150MHz & 202MHz\\
    \hline
    Recompilation Time & N/A & N/A & 3 hr & N/A & 0 hr\\
    \hline
    Winograd & Yes & Yes & No & No & No\\
    \hline
    Run-time Flexibility & No & No & No & No & Yes\\
    \hline
    \end{tabular}
    \label{tab:final results}
 \end{table*}
 
Table \ref{tab:final results} shows the comparison with four prior works \cite{ref11_autoSystolic,ref15_pipecnn,ref16_throughput-opt,ref9} on OpenCL-defined FPGA accelerators for running AlexNet based on the ImageNet dataset \cite{deng2009imagenet} with an input size of 227$\times$227$\times$3. As the source codes of \cite{ref9,ref11_autoSystolic,ref16_throughput-opt} are unavailable, the numbers of these three works used for comparison are quoted from the original papers. As the source codes of PipeCNN \cite{ref15_pipecnn} is available, we implementation PipeCNN with an 8-bit fixed-point data format on the same Intel Arria 10 GX FPGA Development board with the maximum computation parallelism that can be mapped by the tool and measure the inference latency based on this implementation for a fair comparison. As shown in Table \ref{tab:final results}, the PipeCNN \cite{ref15_pipecnn} implementation can only achieve a limited DSP block utilization of 25\% ($vec\_size$=16, $CU\_NUM$=16). In our experiment, we observe that the tool fails to map the design with higher parallelism, mainly because of the large fan-out issue at the local memory buffer interfaces that causes routing congestion.

Our Systolic-CNN accelerator outperforms the prior work in \cite{ref15_pipecnn} and \cite{ref16_throughput-opt} by 6.1x and 5.5x, respectively, in terms of inference latency. It should be noted that the Stratix-V FPGA used in \cite{ref16_throughput-opt}, although running at a lower system frequency, has more logic, memory, and DSP block resources than the Arria 10 FPGA that we use. The prior work in \cite{ref11_autoSystolic} shows a 2.5x better inference latency than our Systolic-CNN accelerator. This is because Winograd transformation \cite{ref33_winograd} is adopted in \cite{ref11_autoSystolic}, which promises to reduce the computational complexity of a convolution layer by a factor of 4x \cite{winogradCNN} to further accelerate CNN inference. \cite{ref9} only tests the on AlexNet, however, the performance of mapping any other models were unknown. For a rough estimation, by introducing 16-bit floating-point, Winograd transformation \cite{winogradCNN} and batch processing mode as \cite{ref9} does, we can improve latency by 2x (estimated), 4x (estimated) and 1.3x (actual), respectively. The total improvement (around 10x) can fill the current gap between \cite{ref9} and ours. While the prior work in \cite{ref11_autoSystolic} fails to fully utilize the available DSP block resource on the FPGA, Systolic-CNN shows better scalability and can achieve up to 100\% utilization of the DSP block resource to fully take advantage of the FPGA device capability. In addition, while the OpenCL kernels of all the prior works are model-specific, Systolic-CNN is invariant to CNN models and have the run-time flexibility needed for handling the dynamic workload of accelerating different CNN models in multi-tenancy cloud/edge computing without the need of the recompilation nor redeployment of the FPGA kernel.   

\begin{table*}[tb]
    \centering
    \caption{{ResNet Inference Comparison of FPGA-based Accelerators with 100\% DSP Resource Utilization.}}
    \begin{tabular}{|p{100pt}|p{70pt}|p{70pt}|p{70pt}|p{40pt}|p{50pt}|}
    \hline
    Work & \cite{ma2018optimizing} & \cite{ma2018optimizing} & \cite{arash19shortcut} & \multicolumn{2}{c|}{This work}\\
    \hline
    CNN Model & ResNet-50 & ResNet-152 & ResNet-152 & ResNet-50 & ResNet-152\\
    \hline
    Data Format & 16-bit fixed & 16-bit fixed & 16-bit fixed & 32-bit float & 32-bit float\\
    \hline
    FPGA & Arria 10 GX1150 & Arria 10 GX1150 & Virtex-7 485T & \multicolumn{2}{c|}{Arria 10 GX1150}\\
    \hline
    Logic Utilization & 221K/427K (52\%) & 235K/427K (55\%) & 372K/433K (86\%) & \multicolumn{2}{c|}{250K/427K (59\%)}\\
    \hline
    Memory Utilization & 1931/2713 (71\%) & 2365/2713 (87\%) & 2039/2060 (99\%) & \multicolumn{2}{c|}{2472/2713 (91\%)}\\
    \hline
    DSP Utilization & 1518/1518 (100\%) & 1518/1518 (100\%) & 2800/2800 (100\%) & \multicolumn{2}{c|}{1518/1518 (100\%)}\\
    \hline
    $f_{CLK}$ & 200 MHz & 200 MHz & 150 MHz & \multicolumn{2}{c|}{202MHz}\\
    \hline
    Inference Latency & 13ms & 32ms & 35ms & 84ms & 202ms \\
    \hline
    Accuracy Degradation & <2\% & <2\% & <1\% & \multicolumn{2}{c|}{0\%} \\
    \hline
    Implementation Method & Verilog & Verilog & C/C++ HLS & \multicolumn{2}{c|}{OpenCL} \\
    \hline
    Winograd & No & No & No & \multicolumn{2}{c|}{No} \\
    \hline
    Recompilation & Yes & Yes & Yes & \multicolumn{2}{c|}{No}\\
    \hline
    \end{tabular}
    \label{tab:table2}
 \end{table*}


The Systolic-CNN results in Table \ref{tab:final results} are measured with the batch processing mode turned on and off (batch size=1). The batch processing mode of Systolic-CNN can efficiently reduce the average latency of fully-connected layer computation. Since AlexNet has intensive computation in the fully-connected layers, one can enable the batch processing mode in Systolic-CNN (batch size=$reuse\_fac$=4) to improve the inference latency of the fully connected layers by 4x, which can further improve the average inference latency of the entire AlexNet by 1.3x. The comparison results in Table \ref{tab:final results} also reflect the overhead of enabling runtime flexibility.

\begin{table*}[tb]
\caption{Inference Performance of Running Different Models on Systolic-CNN Accelerators.}
\centering
\begin{tabular}{|p{65pt}|p{30pt}|p{30pt}|p{30pt}|p{35pt}|p{35pt}|p{30pt}|p{30pt}|p{30pt}|p{35pt}|p{35pt}|}
\hline
FPGA Board & \multicolumn{5}{c|}{Arria 10 GX1150} & \multicolumn{5}{c|}{Stratix 10 GX2800}\\
\hline
Logic Utilization & \multicolumn{5}{c|}{250K/427K (59\%)} & \multicolumn{5}{c|}{562K/933K (60\%)}\\
\hline
Mem. Utilization & \multicolumn{5}{c|}{2472/2713 (91\%)} & \multicolumn{5}{c|}{9611/11721 (82\%)}\\
\hline
DSP Utilization & \multicolumn{5}{c|}{1518/1518 (100\%)} & \multicolumn{5}{c|}{5240/5760 (91\%)}\\
\hline
$f_{CLK}$ & \multicolumn{5}{c|}{200 MHz} & \multicolumn{5}{c|}{172 MHz}\\
\hline
CNN Model & AlexNet & ResNet-50 & ResNet-152 & RetinaNet & LW-RetinaNet & AlexNet & ResNet-50 & ResNet-152 & RetinaNet & LW-RetinaNet  \\
\hline
GFLOPs & 1.4 & 8 & 22 & 312 & 178 & 1.4 & 8 & 22 & 312 & 178 \\
\hline
Latency(ms) & 7 & 84 & 202 & 1615 & 900 & 2 & 33 & 73& 873 & 498\\
\hline
\end{tabular}
\label{tab : retinanet results}
\end{table*}


We also compare the inference performance of running ResNet-50 and ResNet-152 \cite{resnet} with ImageNet dataset (224$\times$224$\times$3) \cite{deng2009imagenet} classification tasks on Systolic-CNN with prior FPGA-based accelerators, as shown in Table \ref{tab:table2}. Here, we mainly compare with two prior works \cite{ma2018optimizing,arash19shortcut} that achieve 100\% DSP resource utilization. \cite{ma2018optimizing} is an RTL-level fine-grained accelerator design with design variables quantitatively investigated, while \cite{arash19shortcut} is more focusing on leverage the off-chip feature map traffic with high-level synthesis (HLS) design flow. There are also other works that design OpenCL-based FPGA accelerator for ResNet models, such as \cite{colangelo}. Since \cite{colangelo} does not provide the latency information nor have open-source code, we do not include it for comparison here. For Systolic-CNN, we use the same kernel as the one used for AlexNet to run the ResNet-50 and ResNet-152 model with no need for recompilation. In terms of the data format and accuracy rate, 32-bit floating-point Systolic-CNN has no accuracy degradation, while the other two works \cite{ma2018optimizing,arash19shortcut} with a 16-bit fixed-point data format can lead up to a 2\% accuracy drop. As CNN grows deeper, it targets more on error-sensitive applications. Systolic-CNN is the one that more suitable for supporting error-sensitive applications in a multi-tenancy cloud/edge computing environment. \cite{ma2018optimizing} performs 6x better than our design in terms of inference latency, which reflects of the performance gap between the two data formats. As 32-bit floating- to fixed-point conversion can introduce ~2.5x speedup \cite{floatingPointClaim} and 32-bit fixed-point to 16-bit fixed-point can offer another ~2x speedup, 5x speedup in total can almost fill the gap of the latency performance between \cite{ma2018optimizing} and ours. At the same time, Systolic-CNN enjoys no recompilation and zero accuracy degradation. \cite{arash19shortcut} also shows 6x better in inference latency than our design. Besides the data format difference between \cite{arash19shortcut} and our work, \cite{arash19shortcut} has ~2x DSP block resources on their FPGA board. Considering both the data format and on-board DSP resource projection, Systolic-CNN performs better than \cite{arash19shortcut} in terms of both latency and accuracy performance.

Table \ref{tab : retinanet results} summarizes the performance of Systolic-CNN accelerator evaluated on five different CNN models -- AlexNet, ResNet-50, ResNet-152, RetinaNet and light-weight RetinaNet (LW-RetinaNet) with Intel Arria 10 and Stratix 10 FPGA, respectively. The evaluation on the same FPGA is done without any recompilation. The inference latency of RetinaNet/LW-RetinaNet is measured based on the COCO dataset \cite{coco} with an input size of 800$\times$800$\times$3 for the object detection task. The DSP block utilization of both implementations is over 90\%, which validates the efficiency of the proposed architecture parameter exploration. In addition, we can see a 2x-3x constant latency improvement between the same model mapped onto two FPGA boards, reflecting the scalability of the proposed Systolic-CNN.


In summary, when mapped with the single-precision floating-point data format, our Systolic-CNN accelerator can achieve an average inference latency of 7ms/2ms, 84ms/33ms, 202ms/73ms, 1615ms/873ms, and 900ms/498ms per image for running AlexNet, ResNet-50, ResNet-152, RetinaNet, and Light-weight RetinaNet on Arria/Stratix 10 FPGA board, respectively. The peak computational throughput is measured at 80-210 GFLOPS/s and 242-700 GFLOPS/s for accelerating different single-precision CNN models on Arria/Stratix 10 FPGA board. Since the current Systolic-CNN architecture is compatible for Winograd-based convolutions, we would like to explore adding support for Winograd-based CNN models to further improve its inference latency performance as future work.


\section{Conclusion}
In this paper, we present Systolic-CNN, an OpenCL-defined scalable, run-time-flexible FPGA accelerator architecture for accelerating CNN inference in cloud/edge computing. Systolic-CNN adopts a highly pipelined and paralleled 1-D systolic array architecture, which efficiently explores both spatial and temporal parallelism for accelerating CNN inference on FPGAs. Systolic-CNN is highly scalable and parameterized with three architectural parameters. By optimizing the architectural parameters, one can efficiently utilize the available DSP block resource to maximize the computational throughput of CNN inference subject to the available off-chip memory bandwidth given an FPGA computing device. The experiment results based on an Intel Arria/Stratix 10 GX FPGA Development board for accelerating five different CNN models validate the scalability and run-time flexibility of Systolic-CNN, which makes it suitable for providing acceleration-as-a-service in cloud/edge computing. The invariance of Systolic-CNN architecture design for CNN models is the key to avoiding the long FPGA compilation time for handling the dynamic workload of accelerating different CNN models in a multi-tenancy cloud/edge computing environment. 



\bibliographystyle{ACM-Reference-Format}

\appendix

\end{document}